\documentstyle[12pt,axodraw]{article}
\begin{document}
\baselineskip 0.6cm
\newcommand{\gsim}{ \mathop{}_{\textstyle \sim}^{\textstyle >} }
\newcommand{\lsim}{ \mathop{}_{\textstyle \sim}^{\textstyle <} }
\newcommand{\vev}[1]{ \left\langle {#1} \right\rangle }
\newcommand{\bra}[1]{ \langle {#1} | }
\newcommand{\ket}[1]{ | {#1} \rangle }
\newcommand{\EV}{ {\rm eV} }
\newcommand{\KEV}{ {\rm keV} }
\newcommand{\MEV}{ {\rm MeV} }
\newcommand{\GEV}{ {\rm GeV} }
\newcommand{\TEV}{ {\rm TeV} }


\begin{titlepage}

\begin{flushright}
UT-883\\
\end{flushright}

\vskip 2cm
\begin{center}
{\large \bf  Quintessence Axion Potential \\}
{\large \bf  Induced by Electroweak Instanton Effects}

\vskip 1.2cm
Yasunori Nomura$^a$, T.~Watari$^a$ and T.~Yanagida$^{a,b}$

\vskip 0.4cm
$^{a}$ {\it Department of Physics, University of Tokyo, \\
         Tokyo 113-0033, Japan}\\
$^{b}$ {\it Research Center for the Early Universe, University of Tokyo,\\
         Tokyo 113-0033, Japan}

\vskip 1.5cm
\abstract{Recent cosmological observations suggest the presence of small 
 but nonzero cosmological constant 
 $\Lambda_{\rm cos}^4 \simeq (2 \times 10^{-3}~\EV)^4$.
 It is an intriguing possibility that such a small cosmological constant 
 is supplied by the potential energy density of an ultralight axion-like 
 field (called as quintessence axion).
 If this axion couples to the electroweak gauge fields, its potential
 may be generated by the electroweak instanton effects.
 We calculate the axion potential assuming the supersymmetric standard
 model and obtain a surprising result that the induced energy density of 
 the quintessence axion field is very close to the value suggested from 
 the observations.}

\end{center}
\end{titlepage}


\section{Introduction}

The problem of cosmological constant is one of the most difficult
problems not only in theory of quantum gravity but also in particle
physics \cite{CC_Review}.
Its observational limit is some 120 orders of magnitude smaller than the 
natural unit of gravity.
On the other hand, one might expect that zero point energies of
dynamical fields would produce a cosmological constant of the order of
the Planck scale.
Even if there was a supersymmetry (SUSY) one would get an induced
cosmological constant of order $m_{\rm SUSY}^4$ whenever the SUSY is
spontaneously broken, where $m_{\rm SUSY}$ is the SUSY-breaking energy
scale.
If one takes $m_{\rm SUSY} \simeq 1~\TEV$ it exceeds the
observational upper limit by 60 orders of magnitude.

A number of attempts have been made to solve this problem, but no
satisfactory solution has been found, so far \cite{CC_Review, Attempt1,
Attempt2, Attempt3, Attempt4, Attempt5, Attempt6, Attempt7}.
However, we expect that some yet unknown mechanism leads to an exactly
vanishing vacuum-energy density ({\it i.e.} zero cosmological constant),
since it seems easier to obtain zero than to obtain a minuscule finite
value.

Contrary to the above expectation, however, there is accumulating
evidence  \cite{Om_lt_1, SN_Ia} that the universe is indeed 
accelerating now.
That is, the recent astrophysical observations strongly indicate the
presence of a small but nonvanishing cosmological constant 
\cite{Om_lt_1, SN_Ia} 
\begin{eqnarray}
  \Lambda_{\rm cos}^4 \simeq (2 \times 10^{-3}~\EV)^4.
\label{vac-en-dens}
\end{eqnarray}
Thus, the situation becomes more puzzling and serious: we must explain
not only why the cosmological constant $\Lambda_{\rm cos}$ is so small
but also why it has the observed magnitude.

We consider that any mechanism leading to the exactly vanishing 
vacuum-energy density may operate only at the true minimum of potential, 
since otherwise the inflation in the early universe does not take place.
This argument has led \cite{ULaxion1, ULaxion2, ULaxion3} to assume 
that the observed cosmological constant is the potential energy density
of a dynamical scalar field ${\cal A}$ which differs from the ground
state value.
Namely, the potential $V({\cal A})$ has a stable minimum at which
$V=0$, but the field ${\cal A}$ has not yet reached the minimum point
since the potential $V$ is very flat.
This requires that the effective mass $m_{\cal A}$ of ${\cal A}$ is not
more than the expansion rate $H_0$ of the present universe:
\begin{eqnarray}
  m_{\cal A} \lsim H_0 \simeq 2 \times 10^{-33}~\EV.
\label{slow-roll}
\end{eqnarray}
The most natural candidate for such an ultralight field ${\cal A}$ is
a pseudo Nambu-Goldstone boson of a nonlinearly realized global
symmetry, which acquires its potential through instanton effects
\cite{ULaxion1}.\footnote{
In the previous paper \cite{previous}, we calculated the potential
$V({\cal A})$ assuming a hypothetical SU($N$) gauge symmetry broken at
the Planck scale.}
There has been noted that such a global symmetry is explicitly broken by 
quantum gravity effects \cite{Coleman}.
However, we postulate throughout this paper that those effects are
sufficiently suppressed, since the dynamics of quantum gravity is not
well known yet.

The purpose of this paper is to calculate the potential of the pseudo
Nambu-Goldstone boson ${\cal A}$ (called as quintessence axion) by using
instantons of the standard electroweak SU(2)$_L$ gauge theory.\footnote{
The contribution to the axion potential from electroweak instanton
effects has been discussed in Refs.~\cite{ULaxion1, Choi, previous}.}
Surprising enough is that the electroweak instanton effects induce a
correct magnitude of the potential energy density ({\it i.e.} the
effective cosmological constant) given in Eq.~(\ref{vac-en-dens}).
We consider that this result is very encouraging for the quintessential
axion hypothesis.

We assume the SUSY standard model throughout this paper, since it is a
fascinating candidate beyond the standard model. 
In the minimal SUSY standard model (MSSM), the electroweak SU(2)$_L$
gauge interaction becomes asymptotic non-free above the SUSY particle
threshold. 
We consider that the MSSM is an effective theory below the Planck scale
and use the cutoff at the reduced Planck scale 
$M_{\rm pl} \simeq 2 \times 10^{18}~\GEV$ to avoid the unphysical
divergence.
The SU(2)$_L$ gauge coupling at the high-energy scale $M_{\rm pl}$ is a
bit larger than that at low energies around the electroweak scale, which
is one of important ingredients that induce the axion potential of
appropriate order of magnitude.

\section{The Quintessence Axion Model}

We introduce the hypothetical axion superfield $\Phi_{\cal A}$ which
couples to the electroweak SU(2)$_L$ field strength $W_\alpha^a$
$(a=1,\cdots,3;\, \alpha=1,2)$ as
\begin{eqnarray}
  {\cal L}_{\rm eff} = 
    \int d^2 \theta \frac{1}{32\pi^2}
    \frac{\Phi_{\cal A}}{F_{\cal A}}
    W^{a \alpha}W_\alpha^a + {\rm h.c.}
\label{axion_superpot}
\end{eqnarray}
other than the MSSM particles, where $F_{\cal A}$ is the decay constant
of the axion.
This axion field $\Phi_{\cal A}$ may arise from spontaneous
breakdown of some global U(1)$_X$ symmetry at high energy of order
$F_{\cal A}$, say the reduced Planck scale $M_{\rm pl}$, or perhaps its
origin can be traced to the string theory \cite{C-W}.
The superpotential Eq.~(\ref{axion_superpot}) possesses a global
U(1)$_X$ symmetry generated by the shift $\Phi_{\cal A} \rightarrow
\Phi_{\cal A} + i F_{\cal A} \delta$, which has an anomaly of the
SU(2)$_L$ gauge interaction.
The quintessence axion field ${\cal A}$ is identified with the imaginary 
part of the lowest component of the superfield $\Phi_{\cal A}$.

In addition to the above global U(1)$_X$, however, there is an
accidental U(1)$_{\rm B+L}$ symmetry in the MSSM which also has an
SU(2)$_L$ gauge anomaly.
There is, thus, a linear combination of U(1)$_X$ and U(1)$_{\rm B+L}$
where the anomaly is canceled out, and the axion field remains exactly
massless due to the presence of such an anomaly-free global symmetry.
Therefore, explicit breaking of $U(1)_{\rm B+L}$ is needed to obtain
the nonzero axion potential.
The required $U(1)_{\rm B+L}$-breaking terms are supplied by
nonrenormalizable operators in the K\"ahler potential and in the
superpotential.\footnote{
We assume that the $R$-parity is exact throughout this paper.}
In this paper, we consider all possible higher-dimensional operators
suppressed by the reduced Planck scale, as long as they are allowed by
symmetries.
Thus, the induced axion potential strongly depends on the symmetries at
the Planck scale.
We impose an U(1)$_R$ symmetry, since it is the most natural symmetry
that guarantees the smallness of vacuum-expectation value (VEV) of the
superpotential $\vev{W} \simeq M_{\rm pl} F$ \cite{IY}, where $F$ is the
SUSY-breaking $F$ term.\footnote{
In any realistic SUSY standard models, the SUSY breaking must occur at
some intermediate scale, $\sqrt{F} \lsim 10^{10}~\GEV$.
Then, to obtain the vanishing cosmological constant at the true minimum
of the potential, the VEV of the superpotential should also be at the
intermediate scale, $\vev{W} \simeq M_{\rm pl} F$, which is naturally
ensured by the approximate $R$-symmetry.}
The U(1)$_R$ charges for the MSSM fields are given in
Table~\ref{R_charge}.
\begin{table}
\begin{center}
\begin{tabular}{|c|ccccc|cc|ccc|}  \hline 
  & $Q$ & $\bar{U}$ & $\bar{D}$ & $L$ & $\bar{E}$ & $H_u$ & $H_d$ &
      ${\rm e}^{-8\pi^2 S}$ & 
      ${\cal D}^2$ & $\bar{\cal D}^2$ \\ \hline
  U(1)$_R$ & $3/5$ & $3/5$ & $1/5$ & $1/5$ & $3/5$ & $4/5$ & $6/5$ &
      $-2$ & $-2$ & $2$ \\ \hline
\end{tabular}
\end{center}
\caption{U(1)$_R$ charges.  
 $Q$, $\bar{U}$, $\bar{D}$, $L$, $\bar{E}$, $H_u$ and $H_d$ represent
 the MSSM chiral superfields.
 $S$ is the gauge-coupling superfield which is related to the SU(2)$_L$
 gauge coupling as in Eq.~(\ref{spurion-S}).
 ${\cal D}^2$ and $\bar{\cal D}^2$ are the super-covariant derivatives.}
\label{R_charge}
\end{table}
All the MSSM interactions are invariant under this U(1)$_R$
symmetry, except for the constant term in the superpotential
$W$.\footnote{
This symmetry is the unique $R$-symmetry which is
consistent with the SU(5)$_{\rm GUT}$ and the neutrino masses generated
through the see-saw mechanism \cite{see-saw}.}
The U(1)$_R$ symmetry restricts possible forms of the ($B+L$)-breaking
nonrenormalizable operators in the K\"ahler potential and in the
superpotential.
In the next section, we estimate the axion potential by introducing
U(1)$_R$ symmetric nonrenormalizable operators.
The breaking of U(1)$_R$ is encoded in the nonzero gaugino mass and the
U(1)$_R$-breaking $A$ terms, $m_{1/2} \simeq A \simeq 1~\TEV$, in the
MSSM sector.

\section{Estimation of the Axion Potential}

The axion potential is generated by the SU(2)$_L$ instantons, which is
given by integrating the size $\rho$ of the instantons.
The effects of the SU(2)$_L$ instantons are screened at distances larger 
than the electroweak scale by nonzero VEV's for the Higgs doublets, and
hence we only have to integrate the region $\rho \lsim \vev{h}^{-1}$.
Thus, we can calculate the axion potential in a manifestly
supersymmetric manner in which all SUSY-breaking effects are described
by VEV's of various superfield spurions \cite{SUSY_axi}.\footnote{
This approximation is, in fact, well justified, since the dominant
contribution to the axion potential comes from the Planck-size
instantons as shown below.}

Let us suppose that the effective Lagrangian at energy scale $\rho^{-1}$ 
is given by
\begin{eqnarray}
  {\cal L} &=& \int d^2\theta d^2\bar{\theta}
    \left[ \sum_r Z_r \Phi_r^{\dagger}\Phi_r 
    +  \sum_i \frac{Y_i}{M_{\rm pl}^{d_i-2}}{\cal O}_i 
    \right] \nonumber\\
  && + \left[ \int d^2\theta \left( \tilde{\mu} H_u H_d
    + \sum_j \frac{\tilde{Y}_j}
    {M_{\rm pl}^{\tilde{d}_j-3}} \tilde{\cal O}_j \right)
    + {\rm h.c.} \right] \nonumber\\
  && + \left[ \int d^2\theta \frac{1}{4} \left( S + 
    \frac{1}{8\pi^2}\frac{\Phi_{\cal A}}{F_{\cal A}} \right) 
    W^{a \alpha}W_\alpha^a + {\rm h.c.} \right],
\label{tree-Lag}
\end{eqnarray}
where operators ${\cal O}_i$ are the general operators composed of
chiral superfields $\Phi_r$ and $\Phi_r^{\dagger}$, and supercovariant
derivatives ${\cal D}^{\alpha}$ and $\bar{\cal D}_{\dot{\alpha}}$, while 
$\tilde{\cal O}_j$ the holomorphic operators composed solely of
$\Phi_r$'s;
$d_i$ and $\tilde{d}_j$ $(d_i, \tilde{d}_j \geq 3)$ are the mass
dimension of the operators  ${\cal O}_i$ and $\tilde{\cal O}_j$,
respectively.
We here consider the wavefunction renormalization and coupling constants 
as superfield spurions,
\begin{eqnarray}
  Z_r &=& 1 - m_r^2 \theta^2 \bar{\theta}^2,
\label{spurion-Z} \\
  Y_i &=& (1 + C_i\theta^2 + C_i^*\bar{\theta}^2 
           + |D_i|^2 \theta^2 \bar{\theta}^2)\, \eta_i, \\
  \tilde{\mu} &=& (1 + B \theta^2)\, \mu, \\
  \tilde{Y}_j &=& (1 + A_j \theta^2)\, \lambda_j, \\
  S &=& \frac{1}{g^2} + i\frac{\Theta}{8\pi^2}
        - \frac{2m_{1/2}}{g^2} \theta^2,
\label{spurion-S}
\end{eqnarray}
where $\eta_i$ and $\lambda_j$ are the coupling constants;
$m_r$, $C_i$, $D_i$, $B$, $A_j$ and $m_{1/2}$ are the soft SUSY-breaking 
masses of order $m_{\rm SUSY} \simeq 1~\TEV$.

The effects of one instanton and one anti-instanton of the SU(2)$_L$ can
be seen by integrating out all the superfields with vanishing VEV's.
It is summarized in the following effective Lagrangian \cite{SUSY_axi}:
\begin{eqnarray}
  {\cal L} &=& \int d^2\theta d^2\bar{\theta}
    \left[ {\rm e}^{-\left(8\pi^2 S+\frac{\Phi_{\cal A}}{F_{\cal A}}\right)}
    K_{\rm eff}(\Phi_r, \Phi_r^{\dagger}, {\cal D}^{\alpha}, 
    \bar{\cal D}_{\dot{\alpha}}, Z_r, Y_i, \tilde{\mu}, \tilde{\mu}^{\dagger}, 
    \tilde{Y}_j, \tilde{Y}_j^{\dagger}, {\rm e}^{S+S^{\dagger}}; \rho)
    + {\rm h.c.} \right] \nonumber\\
  && + \left[ \int d^2\theta 
    {\rm e}^{-\left(8\pi^2 S+\frac{\Phi_{\cal A}}{F_{\cal A}}\right)}
    W_{\rm eff}(\Phi_r, \tilde{\mu}, \tilde{Y}_j; \rho) + {\rm h.c.} \right],
\label{eff-Lag}
\end{eqnarray}
where $\Phi_r$ represents chiral superfields which have nonvanishing
VEV's.
The forms of the $K_{\rm eff}$ and $W_{\rm eff}$ are constrained by the
symmetries present in Eq.~(\ref{tree-Lag}).
The axion potential is obtained by substituting the VEV's for
superfields, $\vev{\Phi_r}$, and spurions in
Eqs.~(\ref{spurion-Z}-\ref{spurion-S}), into Eq.~(\ref{eff-Lag}) and
integrating Grassmann coordinates $\theta$ and $\bar{\theta}$.

To have the full axion potential, we further have to integrate the
instanton size $\rho$.
However, since it requires ($B+L$)-violating nonrenormalizable
interactions to obtain nonzero potential for the axion, the integral may
be divergent at the ultra-violet limit, $\rho \rightarrow 0$.
If it is the case, we use the cutoff at $\rho \simeq M_{\rm pl}^{-1}$ in 
the integral, and then the axion potential is dominated by instantons of
the Planck size. 

Let us first consider the axion potential coming from the effective
superpotential.
Since both $d^2\theta$ and $\exp(-8\pi^2(S+\Phi_{\cal A}/F_{\cal A}))$
carry the U(1)$_R$ charge of $-2$ in the present model, $W_{\rm eff}$
should have the U(1)$_R$ charge of $4$.
Then, the form of $W_{\rm eff}$ is restricted as
\begin{eqnarray}
  W_{\rm eff} = \rho^{-3} 
    \left( \rho^2 \vev{H_u H_d} \right)^2
    \tilde{f} \left( (\rho \tilde{\mu}),\, \frac{\tilde{Y}_j}
    {(\rho M_{\rm pl})^{\tilde{d}_j-3}} \right),
\label{form_sup}
\end{eqnarray}
where $\tilde{f}$ is a holomorphic function which is constrained by the
symmetries in the tree-level Lagrangian.
From this argument only, we can find that the superpotential
contribution to the axion potential $V$,
\begin{eqnarray}
  V = \Lambda_{\cal A}^4 
      \left( 1-\cos\left(\frac{\cal A}{F_{\cal A}}\right) \right),
\end{eqnarray}
is bounded as
\begin{eqnarray}
  \Lambda_{\cal A}^4  &\lsim& 
    {\rm e}^{-\frac{2\pi}{\alpha_2(M_{\rm pl})}}
    \frac{m_{\rm SUSY} \vev{H_u H_d}^2}{M_{\rm pl}} \nonumber\\
  &\simeq& (7 \times 10^{-9}~\EV)^4,
\label{axi-sup}
\end{eqnarray}
where the factor $m_{\rm SUSY}$ is supplied by integrating the Grassmann 
coordinate $\theta^2$ and we have used 
$\alpha_2(M_{\rm pl}) \simeq 1/23$ to give the numerical value.
The energy density obtained in Eq.~(\ref{axi-sup}) is too small to
explain the cosmological constant in Eq.~(\ref{vac-en-dens}) suggested
from the observations.

Incidentally, the above conclusion does not change even if we introduce
some new physics at the intermediate scale, as long as it does not
change the beta-function of the SU(2)$_L$.
Let us extend, for example, the MSSM introducing nonzero
neutrino masses through the see-saw mechanism \cite{see-saw}.
In the region $\rho \lsim M_R^{-1}$, the effective superpotential
contains the Majorana masses $M_R$ for the right-handed neutrinos and
the Yukawa couplings of them to the left-handed neutrinos and the Higgs
field.
Thus, the function $\tilde{f}$ in Eq.~(\ref{form_sup}) has an additional 
argument $(\rho M_R)$ in this region.
On the other hand, in the region $\rho \gsim M_R^{-1}$ the neutrino mass
is described by the effective operator $\tilde{O} = L L H_u H_u / M_R$,
so that the argument of $\tilde{f}$ is given by $1/(\rho M_R)$
in the region $\rho \gsim M_R^{-1}$.
Therefore, the introduction of new physics always brings parameters
smaller than order one as arguments of the function $\tilde{f}$, and
does not enhance the energy scale $\Lambda_{\cal A}$ of the axion
potential.

We next discuss the axion potential generated from the effective
K\"ahler potential.
The U(1)$_R$ charge of $K_{\rm eff}$ should be $2$, and we find that the
dominant contribution to the axion potential may come from the effective
K\"ahler potential of the following form:
\begin{eqnarray}
  K_{\rm eff} = \rho^{-2} 
    \left( \rho \bar{\cal D}^2 \right)
    f \left( Z_r,\, \frac{Y_i}{(\rho M_{\rm pl})^{d_i-2}},\,
    \frac{\tilde{Y}_j} {(\rho M_{\rm pl})^{\tilde{d}_j-3}},\, 
    \frac{\tilde{Y}_j^{\dagger}} {(\rho M_{\rm pl})^{\tilde{d}_j-3}},\, 
    {\rm e}^{S+S^{\dagger}} \right),
\label{form_K}
\end{eqnarray}
where $f$ is some function constrained by the symmetries of the model.
We see that the resulting axion potential is dominated by the
Planck-size instantons ($\rho \simeq M_{\rm pl}^{-1}$) which give
\begin{eqnarray}
  \Lambda_{\cal A}^4 &\simeq&
    {\rm e}^{-\frac{2\pi}{\alpha_2(M_{\rm pl})}}
    C(\eta_i, \lambda_j, \lambda_j^*)\, 
    m_{\rm SUSY}^3 M_{\rm pl} \nonumber\\
  &\simeq& C(\eta_i, \lambda_j, \lambda_j^*)\, 
    (1~\EV)^4,
\label{estimate}
\end{eqnarray}
where $C(\eta_i, \lambda_j, \lambda_j^*)$ is a numerical coefficient
depending on the coupling constants, $\eta_i$, $\lambda_j$ and
$\lambda_j^*$, in the tree-level Lagrangian.
Here, we have assumed the hidden sector SUSY-breaking scenario where the
soft SUSY-breaking terms are nonvanishing at the Planck scale
\cite{hidden}.\footnote{
The SUSY breaking induces the mass of the order of the SUSY-breaking
scale $m_{\rm SUSY} \simeq 1~\TEV$ for SUSY partners of the axion field
${\cal A}$.}
An explicit example of $K_{\rm eff}$ is provided by considering the
dimension-six and dimension-five operators 
${\cal O}=QQ\bar{U}^{\dagger}\bar{E}^{\dagger}$ and 
$\tilde{\cal O}=QQQL$ in the K\"ahler potential and the superpotential,
respectively.
Then, together with the MSSM Yukawa couplings the $K_{\rm eff}$ is
written as
\begin{eqnarray}
  K_{\rm eff} &=& \rho^{-2} \left( \rho \bar{\cal D}^2 \right)
    \left( Z_{\bar{U}} Z_{\bar{E}} \right)^{-1}
    \left( \tilde{Y}_{Q\bar{U}H_u} \tilde{Y}_{L\bar{E}H_d} \right) 
  \nonumber\\
  && \times \left( \frac{Y_{QQ\bar{U}^{\dagger}\bar{E}^{\dagger}}}
    {(\rho M_{\rm pl})^2} \right)
    \left( \frac{\tilde{Y}_{QQQL}} {(\rho M_{\rm pl})} \right)^2
    f \left( {\rm e}^{S+S^{\dagger}-
      \sum_r \frac{T_r}{4\pi^2}\ln{\cal Z}_r} \right),
\label{explicit}
\end{eqnarray}
and we obtain the axion potential in Eq.~(\ref{estimate}).
The potential for the axion field is given by summing up the
contributions from all possible operators, but its order of magnitude is 
determined by the dominant contributions in Eq.~(\ref{estimate}).
Note that the U(1)$_R$ symmetry is crucial to obtain the above result.
If we do not impose it, we obtain a larger axion potential by a factor
of $M_{\rm pl}/m_{\rm SUSY}$.

The numerical value given in Eq.~(\ref{estimate}) seems somewhat large
to explain the cosmological constant in Eq.~(\ref{vac-en-dens}).
However, it strongly depends on the size of various coupling constants
in the coefficient $C$.
In the next section, we argue that the coefficient $C$ is much smaller
than order one in a realistic SUSY standard model and find that in such
a model the resulting cosmological constant is indeed very close to
the value suggested from the observations.

\section{The Axion Potential in a Realistic SUSY Standard Model}

In the explicit example in Eq.~(\ref{explicit}), we have used the
dimension-five operators $\tilde{\cal O}=QQQL$ to provide the required
explicit breaking of U(1)$_{\rm B+L}$.
However, this operator causes dangerous $d=5$ proton decays
\cite{dim-5}.
Thus, the coefficient of this operator should be sufficiently
suppressed to evade too fast proton decay.
The suppression of desirable amount is naturally attained by imposing
the U(1)$_F$ flavor symmetry to the quark and lepton multiplets.
The U(1)$_F$ symmetry also provides realistic quark and lepton mass
matrices through the small breaking parameter 
$\vev{\phi}/M_{\rm pl} \equiv \epsilon$ \cite{FN}.
Therefore, we here impose the U(1)$_F$ symmetry and evaluate the size of 
$C$ in the present model.

We assign the U(1)$_F$ charge consistent with the standard 
SU(5)$_{\rm GUT}$, for simplicity.
The charge assignment is given in Table~\ref{F_charge},
where ${\bf 10}_i \supset \{Q_i,\, \bar{U}_i,\, \bar{E}_i\}$ and 
${\bf 5}^{\star}_i \supset \{\bar{D}_i,\, L_i\}$ are the quark and
lepton chiral superfields embedded in ${\bf 10}$ and 
${\bf 5}^{\star}$ representations of the SU(5)$_{\rm GUT}$,
respectively, and $i=1,\cdots,3$ denotes the generation index.
\begin{table}
\begin{center}
\begin{tabular}{|c|cccccc|cc|c|}  \hline 
  & ${\bf 10}_1$ & ${\bf 10}_2$ & ${\bf 10}_3$ & 
    ${\bf 5}^{\star}_1$ & ${\bf 5}^{\star}_2$ & ${\bf 5}^{\star}_3$ & 
    $H_u$ & $H_d$ & $\vev{\phi}$ \\ \hline
  U(1)$_F$ & $2$ & $1$ & $0$ & $1$ & $0$ & $0$ & $0$ & $0$ & $-1$ \\ \hline
\end{tabular}
\end{center}
\caption{U(1)$_F$ charges.}
\label{F_charge}
\end{table}
This charge assignment is motivated \cite{SY-R} to explain the
observed quark and lepton masses and mixings including the large mixing
angle for the atmospheric neutrino oscillation \cite{SuperK}.
The U(1)$_F$ breaking parameter $\epsilon$ is determined as
$\epsilon \simeq 1/17$ to reproduce the quark and lepton mass matrices
\cite{SY-R}.
It should be noted here that the present model is not necessarily
consistent with the standard SU(5)$_{\rm GUT}$, since the axion
$\Phi_{\cal A}$ is supposed not to couple to QCD gluon fields.
Therefore, we use the SU(5)$_{\rm GUT}$ representations only for a
classification of the quark and lepton multiplets.

The above U(1)$_F$ has an SU(2)$_L$ gauge anomaly, so that the factor 
$\exp(-8\pi^2(S+\Phi_{\cal A}/F_{\cal A}))$ carries the U(1)$_F$ charge
of $10$.
As a result, the coefficient $C(\eta_i, \lambda_j, \lambda_j^*)$ should
be suppressed at least by the tenth powers of $\epsilon$ to match the
U(1)$_F$ quantum numbers.
In the explicit example in Eq.~(\ref{explicit}), we find that $C$ is
suppressed exactly by the tenth powers of $\epsilon$ if we use the
operators ${\cal O}=Q_1Q_2\bar{U}_3^{\dagger}\bar{E}_3^{\dagger}$, 
$\tilde{\cal O}=Q_1Q_1Q_3L_1,\, Q_2Q_2Q_3L_2,\, Q_3\bar{U}_3H_u$ and
$L_3\bar{E}_3H_d$ to close all the fermion zero modes (see
Fig.~\ref{fig_close}).
Thus, the dominant contribution to the axion potential is given by 
\begin{eqnarray}
  \Lambda_{\cal A}^4 &\simeq&
    {\rm e}^{-\frac{2\pi}{\alpha_2(M_{\rm pl})}}
    c\, \epsilon^{10} m_{\rm SUSY}^3 M_{\rm pl} \nonumber\\
  &\simeq& c \left(\frac{\epsilon}{1/17}\right)^{10}
    (1 \times 10^{-3}~\EV)^4,
\label{final_A}
\end{eqnarray}
where $c$ is an order-one constant, assuming that the magnitudes of all
coupling constants are determined solely by the U(1)$_F$ charges.
The resulting value of $\Lambda_{\cal A}^4$ is very close to the value
Eq.~(\ref{vac-en-dens}) suggested from cosmological observations.
This is very encouraging for identifying the axion field ${\cal A}$ with
a quintessence axion.

The axion potential Eq.~(\ref{final_A}) corresponds to the following
axion mass:
\begin{eqnarray}
  m_{\cal A}^2 &\simeq&
    {\rm e}^{-\frac{2\pi}{\alpha_2(M_{\rm pl})}}
    c\, \epsilon^{10} \frac{m_{\rm SUSY}^3 M_{\rm pl}}
    {F_{\cal A}^2} \nonumber\\
  &\simeq& c \left(\frac{\epsilon}{1/17}\right)^{10}
    \left(\frac{2 \times 10^{18}~\GEV}{F_{\cal A}}\right)^2
    (7 \times 10^{-34}~\EV)^2.
\label{final_m}
\end{eqnarray}
This value satisfies the slow-roll condition Eq.~(\ref{slow-roll}) for 
the axion field as long as $F_{\cal A} \gsim 10^{18}~\GEV$.\footnote{
The condition $F_{\cal A} \gsim 10^{18}~\GEV$ does not necessarily need
to hold, if the axion field lies precisely at the maximal point of the
potential, $1-\cos(\vev{\cal A}/F_{\cal A}) \simeq 2$.}
If we set $F_{\cal A} \simeq M_{\rm pl}$, this quintessence axion has an 
extremely small mass of order $10^{-33}~\EV$.

We here comment on the numerical constant $c$.
The constant $c$ contains several ambiguities, but it is not far from of 
order unity.
One possible source of ambiguities comes from loop factors, instanton
measure and integration of the collective coordinate in instanton
calculus, which may make $c$ small.
We give the counting rule for the $\pi^2$ factors in the Appendix.
We find that these factors do not give extremely strong suppression of
$c$; it is, at most, only $(1/\pi^2)^4$ suppression.
Since the axion potential receives contributions from all possible
instanton diagrams, the number of diagrams may partially compensate the
above suppression factors.
Furthermore, there are ambiguities coming from the actual values of
coupling constants and cutoff scale of instanton-size integrations.
These ambiguities are also of a few orders of magnitude, so that it is
reasonable to consider that the numerical constant $c$ in 
Eqs.~(\ref{final_A}, \ref{final_m}) is of order one.

\section{Discussion and Conclusions}

So far, we have calculated the axion potential assuming the particle
content of the MSSM up to the Planck scale.
However, if there exist some superheavy particles below the Planck scale 
$M_{\rm pl}$ carrying nontrivial quantum numbers of the SU(2)$_L$ gauge
symmetry, the SU(2)$_L$ gauge coupling $\alpha_2(M_{\rm pl})$ at the
Planck scale becomes larger and we may obtain larger potential energy of 
the axion.
We now show that it is not the case if the masses for these superheavy
particles are also determined by their U(1)$_F$ charges, and our
conclusion on the induced cosmological constant remains unchanged.

Let us introduce a pair of chiral superfields $\Psi$ and $\bar{\Psi}$
which belong to the vector-like representations of the SU(2)$_L$ and
have U(1)$_F$ charges $q$ and $\bar{q}$ ($q,\bar{q} > 0$), respectively.
Then, these particles have a superheavy mass of order 
$M_{\rm SH} \simeq \epsilon^{q+\bar{q}}M_{\rm pl}$ through the
superpotential $W = M_{\rm SH} \Psi \bar{\Psi}$.
The introduction of the superheavy particles changes the SU(2)$_L$ gauge 
coupling at the Planck scale as
\begin{eqnarray}
  \frac{1}{\alpha_2(M_{\rm pl})} \rightarrow
  \frac{1}{\alpha_2(M_{\rm pl})} - \frac{2T_{\rm SH}}{2\pi}
  \ln \frac{M_{\rm pl}}{M_{\rm SH}},
\end{eqnarray}
where $T_{\rm SH}$ denotes the Dynkin index of the SU(2)$_L$
representation to which $\Psi$ belongs.
This change enhances the instanton factor appearing in the axion
potential as
\begin{eqnarray}
  {\rm e}^{-\frac{2\pi}{\alpha_2(M_{\rm pl})}} \rightarrow
  {\rm e}^{-\frac{2\pi}{\alpha_2(M_{\rm pl})}}
  \left( \frac{M_{\rm pl}}{M_{\rm SH}} \right)^{2T_{\rm SH}}.
\label{change}
\end{eqnarray}
On the other hand, the introduction of the superheavy particles also
leads to additional fermion zero modes around the instanton
configuration.
To close these additional zero modes, we have to use appropriate
operators which contain $\Psi$ and $\bar{\Psi}$ fields.
As a result, numerical factors $C(\eta_i, \lambda_j, \lambda_j^*)$
appearing in the axion potential has an additional suppression factor of
order $\epsilon^{2(q+\bar{q}) T_{\rm SH}}$, compensating exactly the
enhancement in Eq.~(\ref{change}).
The similar argument can also be applied for the case where the
superheavy particle has a Majorana-type mass.
Thus, we find that the existence of the superheavy particles does not
affect the order of magnitude of the axion potential, 
$\Lambda_{\cal A}^4$.

It is interesting to note that the the gauge coupling unification scale
can be raised up to the reduced Planck scale if there exist the
superheavy particles $\Psi_1$ and $\Psi_2$ at the intermediate scale
which belong to the adjoint representations of the SU(3)$_C$ and the
SU(2)$_L$, respectively \cite{SHP}.
Even then, the above argument shows that the observed cosmological
constant Eq.~(\ref{vac-en-dens}) is explained by the energy density of
the quintessence axion, ${\cal A}$, discussed in this paper.

\section*{Acknowledgments}

Y.N. and T.W. thank the Japan Society for the Promotion of Science for
financial support.
This work was partially supported by ``Priority Area: Supersymmetry and
Unified Theory of Elementary Particles (\# 707)'' (T.Y.).

\appendix
\section{Counting $\pi^2$ Factors}

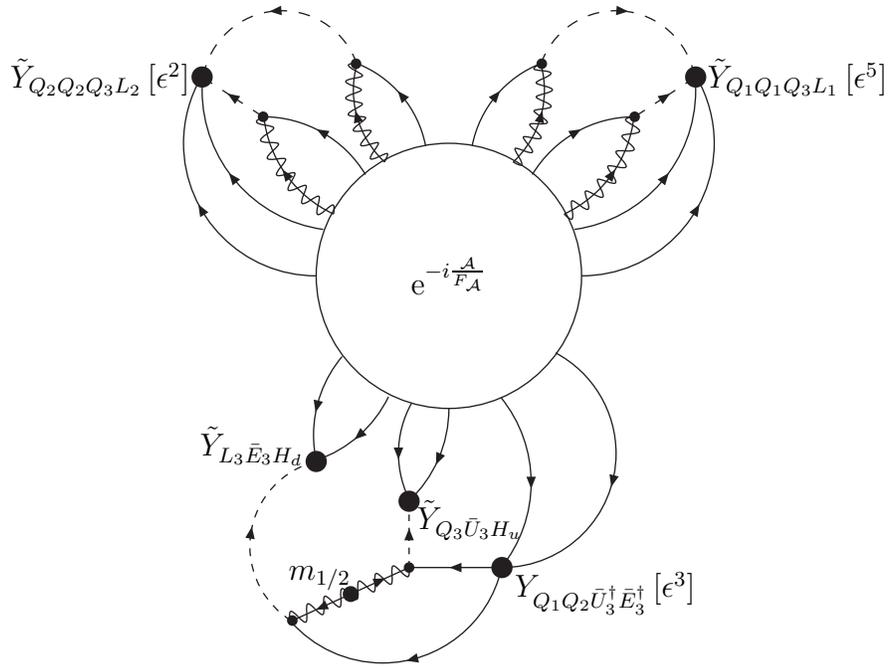
\begin{figure}
\begin{center}
\begin{picture}(300,300)(-150,-150)
\CArc(0,0)(50,0,360)
\Text(0,0)[]{${\rm e}^{-i\frac{\cal A}{F_{\cal A}}}$}
\ArrowArc(50,50)(50,-90,30)               \ArrowArcn(-50,50)(50,270,150)
\ArrowArc(40,70)(53,-82,5)                \ArrowArcn(-40,70)(53,262,175)
\PhotonArc(24,65)(46,-65,-6){3}{7}        \PhotonArc(-24,65)(46,186,245){3}{7}
\ArrowArc(24,65)(46,-65,-6)               \ArrowArcn(-24,65)(46,245,186)
\ArrowArcn(70,15)(45,149,90)              \ArrowArc(-70,15)(45,31,90)
\DashArrowLine(70,60)(93,75){4}           \DashArrowLine(-70,60)(-93,75){4}
\Vertex(70,60){2}                         \Vertex(-70,60){2}
\PhotonArc(-1,70)(37,-47,16){3}{7}        \PhotonArc(1,70)(37,164,227){3}{7}
\ArrowArc(-1,70)(37,-47,16)               \ArrowArcn(1,70)(37,227,164)
\ArrowArcn(50,41)(42,169,111)             \ArrowArc(-50,41)(42,11,69)
\DashArrowArcn(63,70)(30,161,10){4}       \DashArrowArc(-63,70)(30,19,170){4}
\Vertex(35,80){2}                         \Vertex(-35,80){2}
\Vertex(93,75){4}                         \Vertex(-93,75){4}
\Text(98,75)[l]{$\tilde{Y}_{Q_1Q_1Q_3L_1}\,[\epsilon^5]$}
\Text(-98,75)[r]{$\tilde{Y}_{Q_2Q_2Q_3L_2}\,[\epsilon^2]$}
\ArrowArcn(20,-67)(43,62,-90)
\ArrowArcn(-20,-78)(51,39,-39)
\ArrowArcn(-48,-50)(48,0,-47)
\ArrowArc(20,-67)(39,151,209)
\ArrowArcn(-65,-25)(47,-26,-71)
\ArrowArc(-5,-60)(46,140,193)
\DashArrowLine(-15,-110)(-15,-85){3}
\Vertex(-15,-85){4}
\Text(-12,-85)[lt]{$\tilde{Y}_{Q_3\bar{U}_3H_u}$}
\DashArrowArcn(-42,-102)(33,238,104){3}
\Vertex(-50,-70){4}
\Text(-55,-68)[rb]{$\tilde{Y}_{L_3\bar{E}_3H_d}$}
\Photon(-60,-130)(-15,-110){3}{7}
\ArrowLine(-37,-120)(-15,-110)
\ArrowLine(-37,-120)(-59,-130)
\Vertex(-37,-120){3}
\Text(-37,-115)[rb]{$m_{1/2}$}
\ArrowLine(20,-110)(-15,-110)
\ArrowArcn(-25,-100)(46,-13,-139)
\Vertex(-15,-110){2}
\Vertex(-59,-130){2}
\Vertex(20,-110){4}
\Text(25,-112)[lt]{$Y_{Q_1Q_2\bar{U}_3^\dagger\bar{E}_3^\dagger}\,
[\epsilon^3]$}
\end{picture}
\end{center}
\caption{One anti-instanton diagram generating the axion potential, 
${\rm e}^{-i{\cal A}/F_{\cal A}}$.
Together with one instanton diagram which gives 
${\rm e}^{i{\cal A}/F_{\cal A}}$, we obtain the axion potential 
$V = \Lambda_{\cal A}^4 (1-\cos({\cal A}/F_{\cal A}))$ 
(see Ref.~\cite{previous}).}
\label{fig_close}
\end{figure}

One might think at first glance that instanton amplitude is very much
suppressed by loop factors; the diagram given in Fig.~\ref{fig_close},
for example, has many ``loops''. 
However, the meaning of the ``loop'' is not trivial in instanton
calculus, and the naive guess above is not true.
In this appendix, we give a counting rule for $\pi^2$ factors
appearing in instanton diagrams.
The result Eq.~(\ref{eq:pi-counting}) shows that the diagram 
Fig.~\ref{fig_close}, for example, has only $(1/\pi^2)^4$ suppression.

We first count the $\pi^2$ factors analogous to the loop factors in
ordinary perturbation theory.
The instanton calculus are usually formulated in real Euclidean
space-time, where the factors $(1/4\pi^2)$ and $\pi^2$ are provided by
the propagator of nonzero modes and the angular integration of
interaction vertex, respectively.
Thus, we obtain the $\pi^2$ factors,
\begin{eqnarray}
  \left( \frac{1}{\pi^2} \right)^{I-V} = 
  \left( \frac{1}{\pi^2} \right)^{L-C},
\end{eqnarray}
where $I$ and $V$ denote the numbers of internal lines of nonzero modes
and interaction vertices not including the instanton center,
respectively, while $L$ and $C$ are the numbers of loops surrounded only 
by nonzero modes and connected parts after eliminating the instanton
center, respectively.

Another source for the $\pi^2$ factor is the instanton measure. 
The instanton measure consists of collective coordinate measure,
super-product of zero-mode wavefunction norms which comes from the
super-Jacobian of the path-integral functional measure, Pauli-Villars
regularization factor associated with zero modes, and the weight factor
${\rm e}^{-\int d^4x{\cal L}}$ evaluated at the tree level
\cite{NSVZ}.
Among these, the norms of the zero-mode wavefunctions and the
regularization yield $\pi^2$ factors.
The zero-mode wavefunctions are given by the symmetry transformation of
the instanton configuration and the Higgs scalar configuration twisted
by the instanton background which contain no $\pi^2$ factor.
Thus, the zero-mode wavefunctions do not contain $\pi^2$ factor.
However, since the norms of zero-mode wavefunctions are given by
integrating the wavefunctions,
\begin{eqnarray}
   \mbox{norm} = \left( \int d^4x (\mbox{zero-mode wave function})^2 \right)
                 ^{\frac{1}{2}} \propto (\pi^2)^{\frac{1}{2}}.
\end{eqnarray}
they have $\pi^2$ factors.
Since there are $4T_G$ gauge-boson zero modes, $2T_G$ gaugino zero modes
and $2T_R$ matter-fermion zero modes around the instanton configuration,
the super-product of zero-mode wavefunction norms yields 
$(1/\pi^2)^{-(4T_G - 2T_G - 2T_R)/2}$.
Together with the $(1/\sqrt{2\pi})^{4T_G}$ coming from Pauli-Villars
regularization, the total $\pi^2$ factor from the instanton measure is
given by 
\begin{eqnarray}
  \left( \frac{1}{\pi^2} \right)^{T_R}.
\end{eqnarray}

The rest of the sources for the $\pi^2$ factors are integrations of the
collective coordinates. 
Integration of Grassmann collective coordinates does not yield any 
$\pi^2$ factor. 
Integration of the zero modes of gauge-group orientation gives the
volume of coset space SU($N$)$/$SU($N-2$) $\times$ U(1) for SU($N$)
gauge group, which has $(\pi^2)^{N-1}$.
The remaining is only the instanton-size integration.

Combining all factors, we obtain the following instanton amplitude:
\begin{eqnarray}
  \int d^4 x_0  \frac{d \rho^2}{ \rho^2 }  \frac{1}{\rho^{D-b}}
                \left( \frac{1}{\pi^2}\right)^{L-C+T_R-(N-1)}
                \prod ( \mbox{external fields})\,
                \Lambda^{b} {\rm e}^{-4\pi^2 \rho^2 |\vev{h}|^2},
\label{Ap_int}
\end{eqnarray}
where $b = 3T_G-T_R$ and $4-D$ is the sum of the mass dimensions of
external fields in which various coupling constants are included through 
the spurion fields in Eqs.~(\ref{spurion-Z}-\ref{spurion-S}).
We find that the instanton-size integration is always infra-red
convergent because of the exponential factor (Higgs effect) while not
always ultra-violet convergent. 
For the case of $D-b < 0$, the integral Eq.~(\ref{Ap_int}) is fully
convergent and we obtain the $\pi^2$ factor as
\begin{eqnarray}
  \left( \frac{1}{\pi^2} \right)^{L-C+T_R-(N-1)-\frac{1}{2}(D-b)}:
  \qquad \mbox{ for } D-b < 0.
\end{eqnarray}
This reproduces the known result derived in asymptotic-free gauge
theories \cite{NSVZ}.
On the contrary, if $D-b$ is positive, the divergence at the small scale
must be cut off at $\rho_{\rm cut}$, which yields the following
$\pi^2$-counting rule: 
\begin{eqnarray}
  \left( \frac{1}{\pi^2} \right)^{L-C+T_R-(N-1)} 
         \left( \frac{1}{\rho_{\rm cut}M_{\rm pl}} \right)^{D-b}:
  \qquad \mbox{ for } D-b > 0.
\label{eq:pi-counting}
\end{eqnarray}

\newpage
%
%
%
\newcommand{\Journal}[4]{{\sl #1} {\bf #2} {(#3)} {#4}}
\newcommand{\PL}{\sl Phys. Lett.}
\newcommand{\PR}{\sl Phys. Rev.}
\newcommand{\PRL}{\sl Phys. Rev. Lett.}
\newcommand{\NP}{\sl Nucl. Phys.}
\newcommand{\ZP}{\sl Z. Phys.}
\newcommand{\PTP}{\sl Prog. Theor. Phys.}
\newcommand{\NC}{\sl Nuovo Cimento}
\newcommand{\MPL}{\sl Mod. Phys. Lett.}
\newcommand{\PRep}{\sl Phys. Rep.}

%

\newpage
\end{document}